\documentclass[12pt,preprint,apj]{emulateapj}

\newcommand{\kms}{{km s$^{-1}$}}
\newcommand{\kmsMpc}{{km s$^{-1}$ Mpc$^{-1}$}}

\shorttitle{ M66 and M96 distances}
\shortauthors{Lee and Jang}

\begin{document}

\title{
The Tip of the Red Giant Branch Distances to Type Ia Supernova Host Galaxies. II. M66 and M96 in the Leo I Group
}
 
\author{Myung Gyoon Lee and  In Sung Jang}
\affil{Astronomy Program, Department of Physics and Astronomy, Seoul National University, Gwanak-gu, Seoul 151-742, Korea}
\email{mglee@astro.snu.ac.kr, isjang@astro.snu.ac.kr }


\begin{abstract}
M66 and M96 in the Leo I Group are nearby spiral galaxies hosting Type Ia Supernovae (SNe Ia).
We estimate the distances to these galaxies 
from the luminosity of the tip of the red giant branch (TRGB).
We obtain $VI$ photometry of resolved stars in these galaxies from 
$F555W$ and $F814W$ images  in the {\it Hubble Space Telescope} archive.
From the luminosity function of these red giants we find the TRGB $I$-band magnitude 
to be $I_{\rm TRGB}=26.20\pm0.03$ for M66 and  $26.21\pm0.03$ for M96.
These values yield distance modulus  
$(m-M)_0=30.12\pm0.03 ({\rm random})\pm0.12 ({\rm systematic})$ for M66 and
$(m-M)_0=30.15\pm0.03 ({\rm random})\pm0.12 ({\rm systematic})$ for M96.
These results  show that they are indeed the members of the same group.
With these results we derive absolute maximum magnitudes of  two SNe (SN 1989B in M66 and SN 1998bu in M96). 
$V$-band magnitudes of these SNe Ia are $\sim$0.2 mag fainter  than SN 2011fe in M101, the nearest recent SN Ia. 
We also derive near-infrared magnitudes for SN 1998bu.
Optical magnitudes of three SNe Ia (SN 1989B, SN 1998bu, and SN 2011fe) based on TRGB analysis yield a  Hubble constant, $H_0=67.6\pm1.5 ({\rm random})\pm 3.7({\rm systematic})$ \kmsMpc.
This value is similar to the values derived from recent WMAP9 results, $H_0=69.32\pm0.80$ \kmsMpc. 
and from Planck results, $H_0=67.3\pm1.2$ \kmsMpc, 
but smaller than other recent determinations based on Cepheid calibration for SNe Ia luminosity, 
$H_0 = 74 \pm3$ km s$^{-1}$ Mpc$^{-1}$. 
\end{abstract}


\section{Introduction}

Type Ia Supernovae (SNe Ia) are a powerful tool to investigate the expansion history of the universe, because their peak luminosity is as bright as a galaxy and  is known as an excellent standard candle.
Since the discovery of the acceleration of the universe based on the observations of SNe Ia, higher than ever accuracy of their peak luminosity is needed to investigate various problems in cosmology \citep{fre10,rie11,lee12,tam13}. 

We started a project to improve the accuracy of the calibration of the peak luminosity of SNe Ia by measuring accurate distances to nearby resolved galaxies that host SNe Ia. We  derive accurate distances to the SN Ia host galaxies using the method to measure the luminosity of the tip of the red giant branch (TRGB)  \citep{lee93}.
We presented the result of the first target, M101, a well-known spiral galaxy hosting SN 2011fe that is the nearest SN Ia since 1972 (Lee \& Jang 2012 (Paper I)). 
This paper is the second of the series, presenting the results for M66 and M96 in the Leo I Group. 

M66 (NGC 3627, SAB(s)b) and M96 (NGC 3368, SAB(rs)ab) are nearby bright spiral galaxies hosting SNe Ia: SN 1989B in M66 \citep{eva89,wel94} and SN 1998bu in M96 \citep{vil98,sun99,jha99,her00,spy04}. 
M66 has been host to other three SNe as well: SN II 1973R \citep{cia77}, SN imposter SN 1997bs \citep{van00}, and SN II-L 2009hd \citep{eli11}.

They are considered to be the members of the compact Leo I Group that includes three subgroups: the Leo Triplet (M66, M65, and NGC 3628),  the M96 Group (including M96 (NGC 3368), M95 (NGC 3351), and M105 (NGC 3379)), and the NGC 3607 Group \citep{dev75,sah99}. 
The Leo I Group has played an important role as a stepping stone for calibration
of the secondary distance indicators, because it includes both early and late type galaxies at the distance closer than the Virgo cluster and because it hosts SNe Ia.
In particular M66 and M96 have been used as important calibrators for the absolute magnitudes of SNe Ia and  the Tully-Fisher relation \citep{sah99,sun99,sah06,jha07,his11,tam13}.

\citet{har07a} derived a value for the distance to the Leo I Group,
$(m-M)_0\approx 30.10\pm0.05$ ($\approx10.5$ Mpc), from the mean of the known distances to five brightest galaxies in the group (M66, M95, M96, M105, NGC 3351 and NGC 3377). Often the member galaxy candidates without known distances are assumed to be at the same distance, but it is still important to derive a precise distance to each member galaxy candidate for investigating various aspects of the member galaxies.

Unfortunately recent estimates of the distances to M66 and M96 based on resolved stars show a large range \citep{his11,tam13}. 
\citet{sah99} found 68 Cepheids in M66 from  $F555W$ and $F814W$ images obtained with the {\it Hubble Space Telescope (HST)}/Wide Field Planetary Camera 2 (WFPC2) and derived a distance modulus of $(m-M)_0 =  30.22 \pm 0.12$ from the photometry of 25 good Cepheids. 
Later Cepheid estimates range from $(m-M)_0 =  29.70 \pm 0.07$ \citep{wil01} to $ 30.50 \pm 0.09$ \citep{sah06}, showing as much as 0.8 mag differences. 
On the other hand, \citet{mou09a} presented a distance modulus $(m-M)_0 =  29.82 \pm 0.10$ using the TRGB method 
from  $F555W$ and $F814W$ images obtained with the $HST$/Advanced Camera for Surveys (ACS) . Furthermore \citet{tul09} presented an even smaller TRGB distance estimate, $(m-M)_0 =  29.60 \pm 0.09$. 
Thus there  is a significant difference between the Cepheid distances and TRGB distances as well as among the estimates of each method.

In the case of M96, \citet{tan95} found 7 Cepheids from  $HST$/WFPC2 $F555W$ and $F814W$ images 
and derived a distance modulus of $(m-M)_0 =  30.32 \pm 0.16$.
Later Cepheid estimates showed a significant spread, ranging from $(m-M)_0 =  29.94 \pm 0.13$ \citep{wil01} to 
$30.42 \pm 0.15$ \citep{koc97}. 
Surprisingly \citet{mou09b} presented a much smaller TRGB  distance estimate $(m-M)_0 =  29.65 \pm 0.28$ derived from  the $HST$ images.  
Thus the difference between the Cepheid distances and TRGB distance is as much as 0.3 to 0.7 mag and the range of the Cepheid distances is about 0.4.
 
In this study we use the well-known TRGB method to estimate the distances to M66 and M96 from the images  available in the $HST$ archive. 
The TRGB method is an efficient and precise primary distance indicator for resolved galaxies   so that it is an excellent tool for calibration of more powerful distance indicators such as SN Ia and Tully-Fisher relations \citep{lee93,sak96,jan12, tam13}. 
Section 2 describes how we derive photometry of the point sources in the images and \S 3 presents color-magnitude diagrams of the resolved stars in each galaxy, and derive
distances to each galaxy using the TRGB method.
We discuss implications of our results in \S 4, and summarizes
primary results in the final section.

\begin{center}
\begin{table*}[htp]
\centering
\caption{A Summary of $HST$ Observations for M66 and M96}
\begin{tabular}{c c c c c c c}
\hline \hline
Target & R.A. & Dec & Instrument &\multicolumn{2}{c}{Exposure time} & Prop. ID. \\
& (J2000.0) & (J2000.0) & & $F555W$ & $F814W$  \\ 
\hline
 M66 & 11 20 00.00 & 12 59 28.0 & ACS/WFC &   2224 s &  8872 s & 10433 \\
 M96 & 10 46 32.89 & 11 48 16.0 & ACS/WFC &   2280 s &  9112 s & 10433 \\
\hline

\end{tabular}
\end{table*}
\end{center}

\section{Data Reduction}

\begin{figure*}
\centering
\includegraphics[scale=1.0]{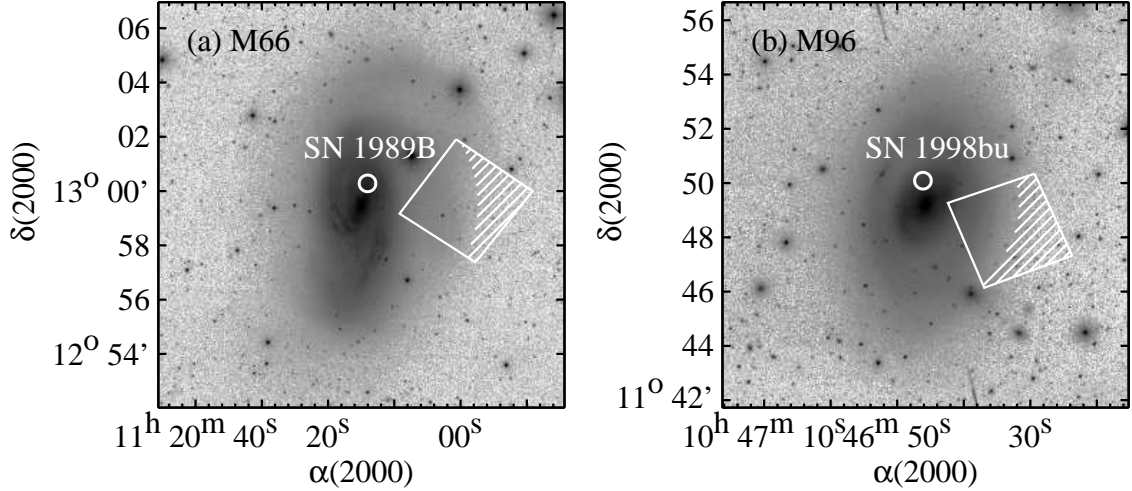} 
\caption{
Finding charts for the $HST$ fields of M66 (a)
and M96 (b) (boxes). Gray scale maps represent  
$i$-band Sloan digital sky survey images.
The hatched regions represent the regions used in the analysis for distance determination. Positions of SN 1989B and SN 1998bu are marked by circles.
}
\label{fig_finder}
\end{figure*}

Table 1 lists the information of the  $HST$/ACS images we used for the TRGB analysis in this study: $F555W$  and $F814W$ images of M66 and M96   (Proposal ID: 10433). 
We made drizzled images for each filter combining the flat fielded images in the HST archive using Tweakreg and AstroDrizzle task in DrizzlePac provided by the Space Telescope Science Institute 
(http://www.stsci.edu/hst/HST$\_$overview/drizzlepac/).
Total exposure times for $F555W$ and $F814W$ are, respectively, 2224 s and 8872 s for M66, and 2280 s and 9112 s for M96. 
In Figure \ref{fig_finder}  we illustrate the locations of the $HST$ fields in the gray scale maps of $i$-band Sloan Digital Sky Survey images of M66 and M96.
The $HST$ fields cover the west region of each galaxy off from the galaxy center. 
Two known SNe Ia (SN 1989B and SN 1998bu) are located close to the center of each galaxy and are not covered by these images, as marked in Figure \ref{fig_finder}. 

Instrumental magnitudes of point sources in the images were obtained using 
the DAOPHOT package in IRAF 
 \citep{ste94}, as done for M101 in \citet{lee12}. Details are described in \citet{lee12}.
Mean values for the aperture correction errors  are 0.02 mag for both filters.
The instrumental magnitudes  were converted into the standard Johnson-Cousins $VI$ magnitudes, using the information
in \citet{sir05}. The average errors for this transformation are 0.02 mag.
We adopted the  standard Johnson-Cousins $VI$ magnitudes for transformation to compare our results with others in the literature and  combine our results with those for other galaxies sometimes based on F606W images. 


\section{Results}

\subsection{Photometry of Resolved Stars} 

The $HST$/ACS fields cover disk regions with spiral arms in each galaxy. We need to select resolved old red giants for the analysis of the TRGB method.  Therefore we selected an outer region avoiding arms in each field, as marked by the hatched region in Figure \ref{fig_finder}. Thus chosen regions have the lowest sky background level in the images. 

\begin{figure*}
\centering
\includegraphics[scale=1.0]{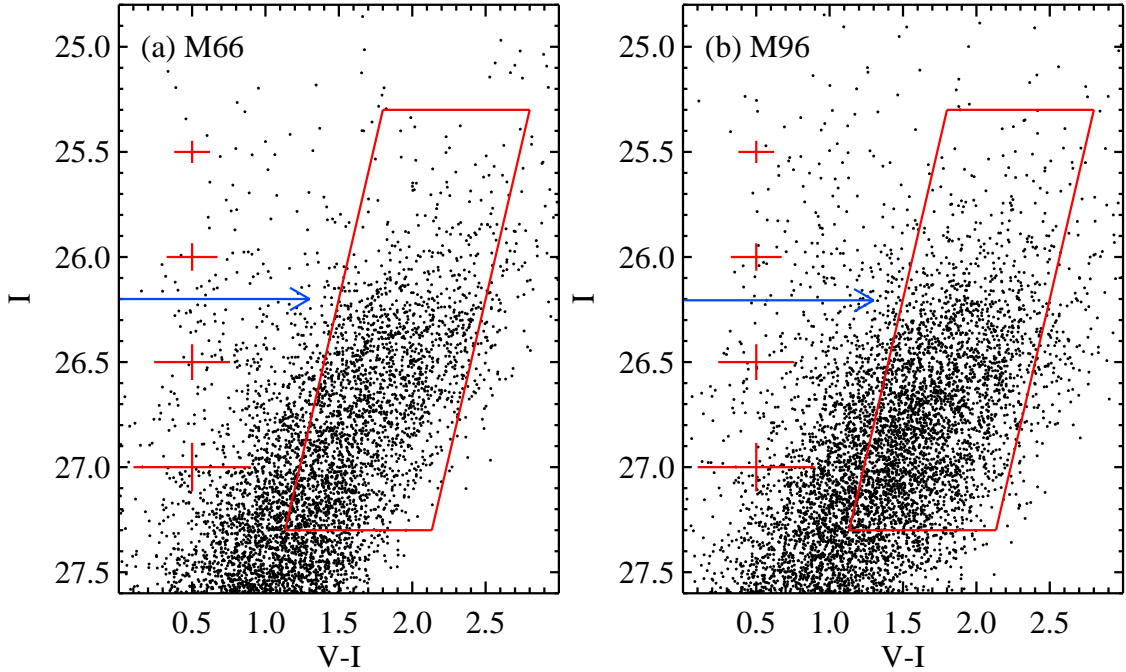} 
\caption{
$I-(V-I)$ color-magnitude diagrams of the  detected stars in the selected regions of M66 (a) and M96 (b).
Boxes denote the boundary of the red giants used 
for distance determination.
Arrows indicate the magnitudes of the TRGB.
 Mean photometric errors for given magnitude bins are
 plotted by error bars.}
\label{fig_cmd}
\end{figure*}

Color-magnitude diagrams (CMDs) of the resolved stars in the selected regions in M66 and M96 are plotted in  Figure \ref{fig_cmd}. It shows that most of the resolved stars in each galaxy are red giants belonging to the thick slanted feature, which is a red giant branch (RGB). The brightest part of the RGB 
is seen at $I \approx 26.2$ mag in each galaxy, which corresponds to the TRGB.
We adopted the foreground reddening values, $E(B-V)=0.028$ for M66 and 0.022 for M96 in \citet{sch98,sch11}. 
These values yield $A_I=0.049$ and $E(V-I)=0.040$ for M66 and  $A_I=0.038$ and $E(V-I)=0.031$ for M96.
We assumed that internal reddening for the old red giants is zero.

\subsection{TRGB Distance Measurement} 
 
\begin{figure*}
\centering
\includegraphics[scale=1.0]{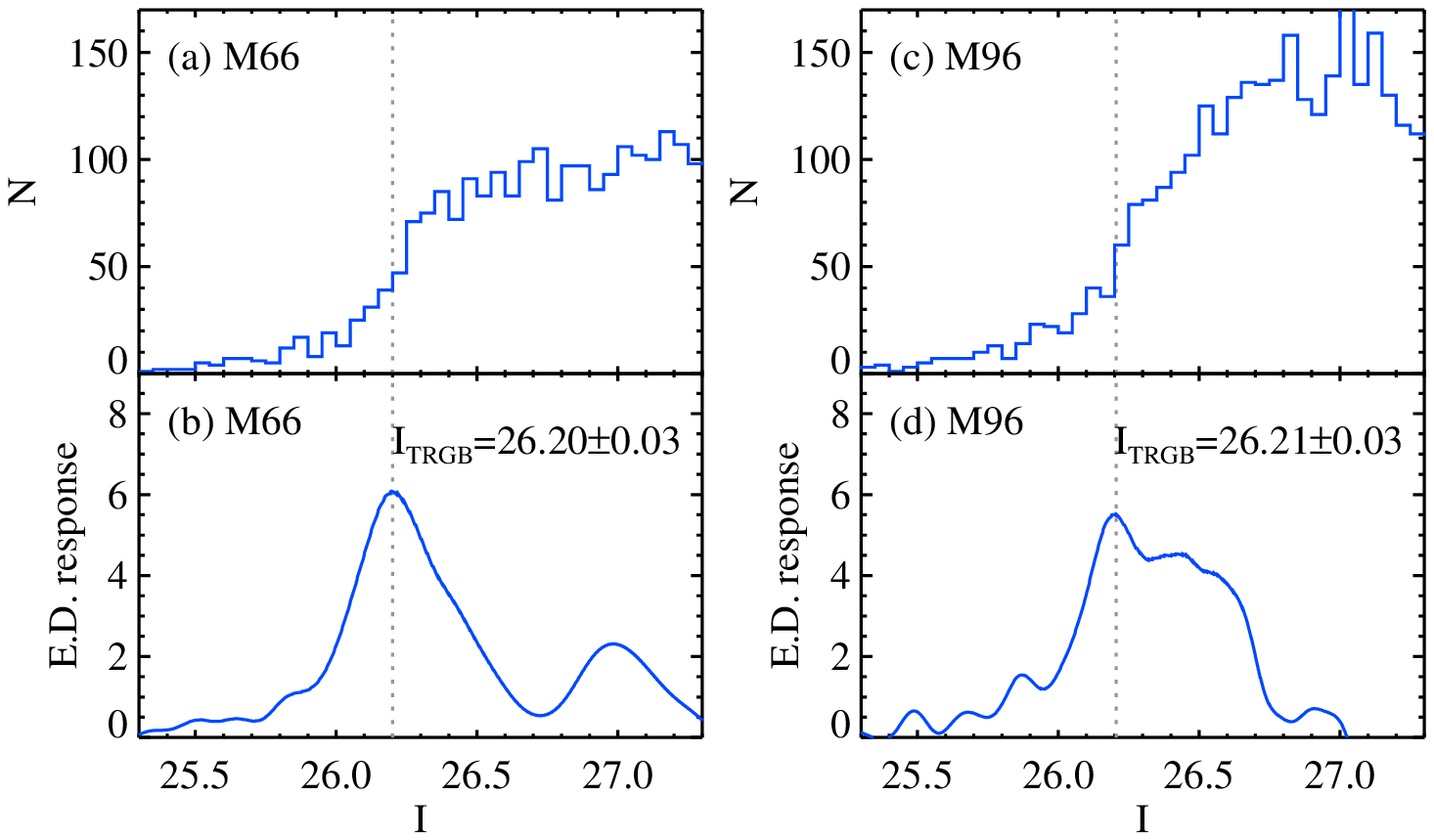} 
\caption{(a) and (c) denote $I$-band luminosity functions of the red giants in the selected regions of M66 and M96, respectively.  (b) and (d)  plot corresponding  edge-detection responses ($E(I)$) for M66 and M96, respectively.  
Note that (b) and (d) show clearly a dominant single peak for each galaxy at the magnitude corresponding to the TRGB position (dotted lines).
}
\label{fig_ilf}
\end{figure*}

We  estimated the distances to M66 and M96  from the photometry of the resolved stars using the TRGB method, as described in \citet{lee12}. 
Figure \ref{fig_ilf}(a) and (c) 
plot the $I$-band luminosity functions of the red giants obtained counting the stars 
inside the box as marked in Figure \ref{fig_cmd}.
In Figure \ref{fig_ilf} an abrupt discontinuity  is seen at $I \approx 26.2$ mag for each galaxy, which is also noticed in the CMDs. 
This  matches the  TRGB in each galaxy.

We performed a quantitative analysis of the TRGB measurement  
using the edge-detecting algorithm \citep{sak96,men02,mou10}.
When the $I$-band luminosity function of the stars is given by $N(I)$  and 
$\sigma_I$ is the mean photometric error,  
the edge-detection response function is given by 
$E(I)$ ($= N (I+ \sigma_I ) -   N (I - \sigma_I )  $). 
The values of the TRGB magnitudes were determined from the peak values of the edge-detection response function.
Figure \ref{fig_ilf}(b) and (d) illustrate the edge-detection response functions for
M66 and M96, respectively. 
The edge-detection response function for each galaxy shows a strong peak at the position corresponding to  the TRGB.
We estimated the measurement errors for the TRGB magnitudes 
using bootstrap resampling method as described in \citet{lee12}.
Thus estimated TRGB magnitudes are $I_{\rm TRGB} = 26.20\pm0.03$ for M66 and $26.21\pm0.03$ for M96, both of
which are almost the same.
We obtained a median color value of the TRGB  from the color of the brightest part of the RGB: 
 $(V-I)_{\rm TRGB} = 1.97\pm0.05$ for M66 and $1.93\pm0.04$ for M96.
For calculating distance moduli from apparent TRGB magnitudes we adopted a relation 
\citet{riz07} derived:
$M_{{\rm I,TRGB}} = -4.05(\pm0.02) + 0.217(\pm0.01)( (V-I)_{0, TRGB} -1.6)$.
After correction for foreground reddening, we derived distance modulus :
$(m-M)_0=30.12\pm0.03$ for M66 and $(m-M)_0=30.15\pm0.03$ for M96 (where 0.03 is a measurement error).
We derived a value of the systematic error  to be 0.12, from the combination of
the TRGB magnitude error, aperture correction error, and standard transformation error, as described in \citet{lee12}. 
Thus derived distance to these galaxies are $10.57\pm0.15\pm0.58$ Mpc for M66 and $10.72\pm0.15\pm0.59$ Mpc for M96.
Our distance estimates for M66 and M96 are summarized in Table 2. 

\begin{table*}[htp]
\centering
\small
\caption{A Summary of TRGB Distance Measurements for M66 and M96}
\begin{tabular}{l c c c}
\hline \hline
Parameter	 & M66 & M96  \\
\hline
TRGB magnitude, $I_{TRGB}$ 					& $26.20\pm0.03$  	& $26.21\pm0.03$ 	& \\
TRGB color,  $(V-I)_{TRGB}$ 				& $2.01\pm0.05$		& $1.96\pm0.04$   	&\\
Foreground extinction at $V$, $A_{V}$ 		& $0.089$ 			& $0.069$  			&\\
Foreground extinction at $I$, $A_{I}$ 		& $0.049$ 			& $0.038$  			&\\
Foreground reddening,  $E(V-I)$ 			& $0.040$ 			& $0.031$   		&\\
Intrinsic TRGB magnitude $I_{0,TRGB}$ 		& $26.15\pm0.03$  	& $26.17\pm0.03$   	&\\
 Intrinsic TRGB color, $(V-I)_{0,TRGB}$ 	& $1.97\pm0.05$  	& $1.93\pm0.04$   	&\\
 Absolute TRGB magnitude, $M_{I,TRGB}$ 		& $-3.97\pm0.12$  	& $-3.98\pm0.12$	&\\
 Distance modulus, $(m-M)_0$ 				& $30.12\pm0.03\pm0.12$  & $30.15\pm0.03\pm0.12$   &\\
 Distance  				& $10.57\pm0.15\pm0.58$  & $10.72\pm0.15\pm0.59$   &\\
\hline
\end{tabular}
\end{table*}

\section{Discussion}

\subsection{Comparison with Previous Distance Estimates}

\begin{figure*}
\centering
\includegraphics[scale=1.1]{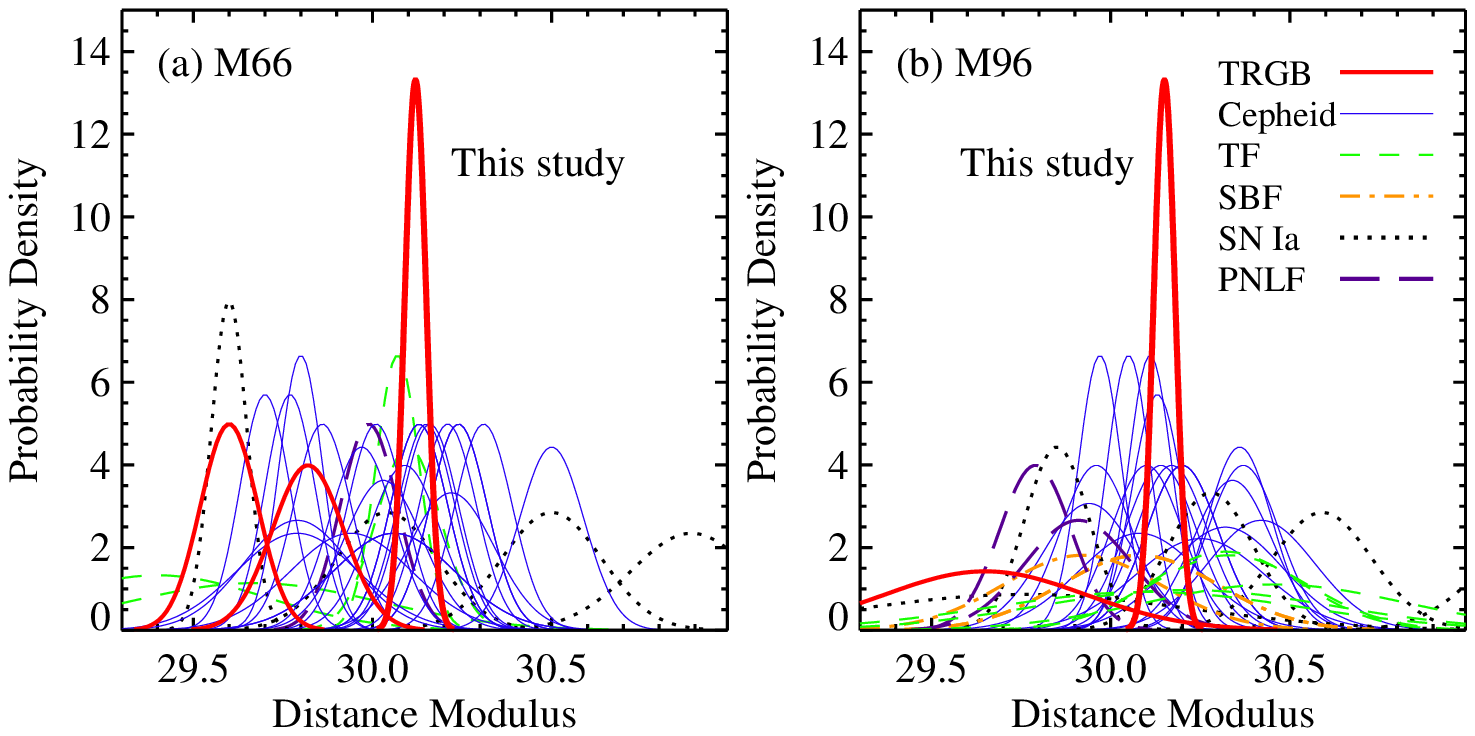} 
\caption{Comparison of the distance measurements for M66 (a) and M96 (b)
 derived in this study and previous studies based on the 
 TRGB (thick solid lines), 
 Cepheids (thin solid lines),
 Tully-Fisher relations (dashed lines),
 SBF (dot-dashed lines),
 SN Ia (dotted lines) and PNLF (long-dashed lines).
%
A probability density curve for each measurement was derived from a Gaussian function centered at the distance modulus value with a width equal to the measurement error. } 
\label{fig_comp}
\end{figure*}

There are numerous previous estimates for the distances to M66 and M96 
based on various methods (TRGB, Cepheids, Tully-Fisher relations, surface brightness fluctuation (SBF), planetary nebula luminosity functions (PNLFs), and SNe Ia), as listed in Tables 3 and 4.
We compare our estimates for the distances to M66 and M96 with these previous estimates. 
Figure \ref{fig_comp} shows a comparison of distance measurements for each galaxy in this study and previous studies. 
We derived a probability density curve for each measurement with a normalized Gaussian function centered at the distance modulus value with a width equal to the measurement error. 

Comparison of the TRGB distances derived in this study and previous
studies \citep{mou09a,tul09} shows significant differences.
Our distance estimate for M66 is 0.3 mag larger than that of \citet{mou09a} 
($(m-M)_0=29.82\pm0.10$) and 0.5 mag larger than that of \citet{tul09} ($(m-M)_0=29.60\pm0.09$). 
In the case of M96, our distance estimate  is 0.5 mag larger than that of \citet{mou09b} 
($(m-M)_0=29.65\pm0.18$). 
These differences are explained in terms of the TRGB magnitude differences: the two previous studies derived much brighter magnitudes for the TRGB than this study. 
\citet{mou09a} and \citet{tul09} presented $I_{\rm TRGB} = 25.83$ and 25.56, respectively, for M66,
which are, respectively, 0.37 mag and 0.64 mag brighter than the our value. 
Similarly \citet{mou09b}  presented $I_{\rm TRGB} = 25.66$ for M96,
which is 0.55 mag brighter than our value. 
What caused these differences is not clear, but the previous measurements might have been affected by younger stars in the disk of each galaxy. Note that we used only the stars in the arm-free regions in each galaxy to reduce the contamination due to younger stars for our analysis. 

Our distance estimate is consistent with some of the previous estimates based on 
other distance indicators (Cepheids, Tully-Fisher relations, SBF, and SN Ia). 
However, the spread in the previous measurements for each method is significant and the errors for each measurement are mostly larger than ours. 
It is expected that our results will be useful for improving the calibration of these other distance indicators in the future.


\begin{table*}[htp]
\centering
\small
\caption{A List of Distance Measurements for M66}
\begin{tabular}{cllcl}
\hline \hline


ID	 & Reference & Method & Distance Modulus & Remarks\\
\hline

\startdata
1 & \citet{pie94} 			& Tully-Fisher	& 29.40	$\pm$ 0.30 & \\
2 & \citet{rus02}			& Tully-Fisher	& 30.10 $\pm$ 0.09 & I band calibration \\
3 &  			 			& Tully-Fisher	& 30.07 $\pm$ 0.06 & B band calibration \\
4 & \citet{tul09}$^a$		& Tully-Fisher	& 29.67	$\pm$ 0.35 & \\
5 & \citet{cia02}			& PNLF$^b$		& 29.99	$\pm$ 0.08 & N(PNe)=40\\
6 & \citet{mue94}			& SN Ia	(Opt)	& 29.60	$\pm$ 0.05 & \\
7 & \citet{rei05}			& SN Ia	(Opt)	& 30.50	$\pm$ 0.14 & $H_0=60.0$ \kmsMpc \\	
8 & \citet{jha07}			& SN Ia	(Opt)	& 30.04	$\pm$ 0.14 & $H_0=65.0$ \kmsMpc \\
9 & \citet{tak08}			& SN Ia	(Opt)	& 30.89	$\pm$ 0.17 & $H_0=70.8$ \kmsMpc \\
10& \citet{sah99}			& Cepheids (LMC)	& 30.22 $\pm$ 0.12 & N(Cep)=25, $(m-M)_{\rm 0,LMC}=18.50$ \\
11& \citet{gib00}			& Cepheids (LMC)	& 30.15 $\pm$ 0.08 & N(Cep)=21, $(m-M)_{\rm 0,LMC}=18.50$ \\
12& 						& Cepheids (LMC)	& 30.06 $\pm$ 0.17 & N(Cep)=17, $(m-M)_{\rm 0,LMC}=18.50$ \\
13& \citet{fre01}			& Cepheids (LMC)	& 30.01 $\pm$ 0.15 & N(Cep)=16, $(m-M)_{\rm 0,LMC}=18.50$ \\
14& 						& Cepheids (LMC)	& 29.86 $\pm$ 0.08 & N(Cep)=16, $(m-M)_{\rm 0,LMC}=18.22$ \\
15& 						& Cepheids (LMC)	& 29.88 $\pm$ 0.08 & N(Cep)=35, $(m-M)_{\rm 0,LMC}=18.50$ \\
16& 						& Cepheids (LMC)	& 29.71 $\pm$ 0.08 & N(Cep)=35, $(m-M)_{\rm 0,LMC}=18.22$ \\
17& \citet{gib01}			& Cepheids (LMC)	& 29.94 $\pm$ 0.17 & N(Cep)=17, $(m-M)_{\rm 0,LMC}=18.45$ \\
18& 						& Cepheids (LMC)	& 29.79 $\pm$ 0.17 & N(Cep)=17, $(m-M)_{\rm 0,LMC}=18.45$ \\
19& \citet{wil01}			& Cepheids (LMC)	& 29.70 $\pm$ 0.07 & N(Cep)=36, $(m-M)_{\rm 0,LMC}=18.50$ \\
20& \citet{dol02}			& Cepheids (LMC)	& 30.09 $\pm$ 0.10 & N(Cep)=28, $(m-M)_{\rm 0,LMC}=18.50$ \\
21& 						& Cepheids (LMC)	& 30.03 $\pm$ 0.11 & N(Cep)=28, $(m-M)_{\rm 0,LMC}=18.50$ \\
22& 						& Cepheids (LMC)	& 29.97 $\pm$ 0.09 & N(Cep)=28, $(m-M)_{\rm 0,LMC}=18.50$ \\
23& \citet{pat02}			& Cepheids (MW)		& 29.80 $\pm$ 0.06 & N(Cep)=25 \\
24& 						& Cepheids (MW)		& 29.77 $\pm$ 0.07 & N(Cep)=25 \\
25& \citet{kan03}			& Cepheids (MW)		& 30.31 $\pm$ 0.08 & N(Cep)=25 \\
26&  						& Cepheids (MW)		& 30.24 $\pm$ 0.08 & N(Cep)=25 \\
27& 						& Cepheids (MW)		& 30.21 $\pm$ 0.08 & N(Cep)=25 \\
28&  						& Cepheids (LMC)	& 30.24 $\pm$ 0.08 & N(Cep)=25, $(m-M)_{\rm 0,LMC}=18.50$ \\
29&  						& Cepheids (LMC)	& 30.16 $\pm$ 0.08 & N(Cep)=25, $(m-M)_{\rm 0,LMC}=18.50$ \\
30&  						& Cepheids (LMC)	& 30.13 $\pm$ 0.08 & N(Cep)=25, $(m-M)_{\rm 0,LMC}=18.50$ \\
31&  						& Cepheids (LMC)	& 30.13 $\pm$ 0.08 & N(Cep)=25, $(m-M)_{\rm 0,LMC}=18.50$ \\
32& \citet{sah06}			& Cepheids (LMC)	& 30.50 $\pm$ 0.09 & N(Cep)=22, $(m-M)_{\rm 0,LMC}=18.50$ \\
33& \citet{tul09}$^a$		& TRGB				& 29.60	$\pm$ 0.09 & $I_{TRGB}=25.56, M_{\rm I,TRGB}=-4.10$\\
34& \citet{mou09a}			& TRGB				& 29.82	$\pm$ 0.10 & $I_{TRGB}=25.83, M_{\rm I,TRGB}=-4.05$\\
35& This study				& TRGB				& 30.12	$\pm$ 0.03 & $I_{TRGB}=26.20, M_{\rm I,TRGB}=-3.97$\\

\hline
\multicolumn{4}{l}{%
  \begin{minipage}{10cm}%
    $^a$ The Extragalactic Distance Database (EDD) \citep{tul09}.\\
    $^b$ The Planetary Nebula Luminosity Function (PNLF).\\

  \end{minipage}%
}\\
\end{tabular}
\end{table*}



\begin{table*}[htp]
\centering
\small
\caption{A List of Distance Measurements for M96}
\begin{tabular}{cllcl}
\hline \hline


ID	 & Reference & Method & Distance Modulus & Remarks\\
\hline

\startdata
1 & \citet{rus02}			& Tully-Fisher		& 30.32$\pm$ 0.21 & B band calibration \\
2 & 						& Tully-Fisher		& 30.33$\pm$ 0.22 & I band calibration \\
3 & \citet{spr09}			& Tully-Fisher		& 30.10$\pm$ 0.43 & \\
4 & 						& Tully-Fisher		& 30.21$\pm$ 0.41 & \\
5 & \citet{tul09}$^a$		& Tully-Fisher		& 30.46$\pm$ 0.36 & \\
6 & \citet{fel97}			& PNLF$^b$			& 29.91$\pm$ 0.15 & N(PNe)=74\\
7 & \citet{cia02}			& PNLF$^b$			& 29.79$\pm$ 0.10 & N(PNe)=74\\
8 & \citet{rei05}			& SN Ia	(Opt)		& 30.59$\pm$ 0.14 & $H_0=60.0$ \kmsMpc \\
9 & \citet{jha07}			& SN Ia	(Opt)		& 30.28$\pm$ 0.12 & $H_0=65.0$ \kmsMpc \\
10 & \citet{tak08}			& SN Ia	(Opt)		& 31.20$\pm$ 0.17 & $H_0=70.8$ \kmsMpc \\
11& \citet{woo08}			& SN Ia (NIR)		& 29.76$\pm$ 0.46 & $H_0=72.0$ \kmsMpc  \\
12& \citet{man09}			& SN Ia (NIR)		& 29.85$\pm$ 0.09 & $H_0=72.0$ \kmsMpc  \\
13& \citet{ajh01}			& SBF$^c$			& 30.08$\pm$ 0.22 &    \\
14& \citet{ton01}			& SBF$^c$			& 30.08$\pm$ 0.22 &    \\
15& \citet{jen03}			& SBF$^c$			& 29.92$\pm$ 0.22 &  \\
16& \citet{tan95}			& Cepheids (LMC)	& 30.32$\pm$ 0.16 & N(Cep)=7, $(m-M)_{\rm 0,LMC}=18.50$\\
17& \citet{koc97}			& Cepheids (LMC)	& 30.14$\pm$ 0.10 & N(Cep)=7, $(m-M)_{\rm 0,LMC}=18.50$\\
18& 						& Cepheids (LMC)	& 30.42$\pm$ 0.15 & N(Cep)=7, $(m-M)_{\rm 0,LMC}=18.50$\\
19& \citet{tan99}			& Cepheids (LMC)	& 30.13$\pm$ 0.07 & N(Cep)=16, $(m-M)_{\rm 0,LMC}=18.50$\\
20& 						& Cepheids (LMC)	& 30.25$\pm$ 0.18 & N(Cep)=16, $(m-M)_{\rm 0,LMC}=18.50$\\
21& \citet{kel00}			& Cepheids (LMC)	& 30.37$\pm$ 0.10 & N(Cep)=7, $(m-M)_{\rm 0,LMC}=18.50$\\
22& \citet{gib00}			& Cepheids (LMC)	& 30.20$\pm$ 0.10 & N(Cep)=7, $(m-M)_{\rm 0,LMC}=18.50$\\
23& 						& Cepheids (LMC)	& 30.36$\pm$ 0.09 & N(Cep)=7, $(m-M)_{\rm 0,LMC}=18.50$\\
24& \citet{wil01}			& Cepheids (LMC)	& 29.94$\pm$ 0.13 & N(Cep)=11, $(m-M)_{\rm 0,LMC}=18.50$\\
25& \citet{gib01}			& Cepheids (LMC)	& 29.96$\pm$ 0.10 & N(Cep)=7, $(m-M)_{\rm 0,LMC}=18.50$\\
26& 						& Cepheids (LMC)	& 30.10$\pm$ 0.10 & N(Cep)=7, $(m-M)_{\rm 0,LMC}=18.50$\\
27& \citet{fre01}			& Cepheids (LMC)	& 29.97$\pm$ 0.06 & N(Cep)=9, $(m-M)_{\rm 0,LMC}=18.22$\\
28& 						& Cepheids (LMC)	& 30.11$\pm$ 0.06 & N(Cep)=9, $(m-M)_{\rm 0,LMC}=18.50$\\
29& 						& Cepheids (LMC)	& 29.95$\pm$ 0.08 & N(Cep)=11, $(m-M)_{\rm 0,LMC}=18.22$\\
30& 						& Cepheids (LMC)	& 30.10$\pm$ 0.08 & N(Cep)=11, $(m-M)_{\rm 0,LMC}=18.50$\\
31& \citet{pat02}			& Cepheids (MW)		& 30.05$\pm$ 0.06 & N(Cep)=7 \\
32& 						& Cepheids (MW)		& 30.17$\pm$ 0.10 & N(Cep)=7 \\
33& \citet{sah06}			& Cepheids (LMC)	& 30.34$\pm$ 0.11 & N(Cep)=7, $(m-M)_{\rm 0,LMC}=18.54$\\
34& \citet{mou09b}			& TRGB			& 29.65$\pm$ 0.28 &$ I_{TRGB}=25.66, M_{\rm I,TRGB}=-4.04$\\
35& This study				& TRGB			& 30.15$\pm$ 0.03 &$ I_{TRGB}=26.21, M_{\rm I,TRGB}=-3.98$\\
\hline
\multicolumn{4}{l}{%
  \begin{minipage}{10cm}%
    $^a$ The Extragalactic Distance Database (EDD) \citep{tul09}.\\
    $^b$ The Planetary Nebula Luminosity Function (PNLF).\\
    $^c$ The Surface Brightness Fluxuation (SBF).\\

  \end{minipage}%
}\\
\end{tabular}
\end{table*}

\subsection{The Membership of the Leo I Group}

The distance estimates derived in this study show that M66 and M96 are at the same distance and that they are located at the same distance as
the mean distance to the Leo I Group \citep{har07a}. 
These confirm that M66 and M96 are indeed the members of the Leo I Group.

M96 is the brightest member of the Leo I Group. 
However it is not located at the center of the M96 Group.  An E1 galaxy M105 resides at the center of the M96 Group, and M96 is 48$\arcmin$ at the south-west of the group center.  M96 has a large pseudo bulge \citep{now10}, and appears to be connected to a tidal feature extended out from the well-known giant HI ring surrounding a pair of M105 and NGC 3384 (SB0) \citep{sch83,sch89}. Whether this giant gas ring around M105/NGC 3384 is primordial or formed via collision of disk galaxies (M105/NGC 3384 and M96) has been controversial \citep{thi09,mic10}. 
Precise distance estimates of M96 and M105/NGC 3384 will be useful to investigate the origin 
of this giant ring, because the relative distances (as well as velocities) are critical constraints for simulation models \citep{mic10}.

Here we compare the distance to M96 with that of M105.
\citet{har07b} estimated the $I$-band magnitude of the TRGB
for M105 to be $I_{TRGB}=26.10\pm0.10$ from the $HST$/ACS $F606W$ and $F814W$ images of a field $630\arcsec$ west and $173\arcsec$ north of the galaxy center, 
and derived a distance modulus  $(m-M)_0=30.10\pm0.16$ adopting the foreground reddening
of $A_I=0.05\pm0.02$ and the absolute TRGB magnitude given in 
\citet{bel04}, $M_{I,TRGB}=-4.05\pm0.12$.
This value is nearly the same as the TRGB distance to M96 derived in this study,
showing that M105 and M96 are at the same distance.
The radial velocities of M96 and M105 are
also very similar ( $897\pm 4$ \kms  and $911\pm 2$ \kms, respectively), while they are $\sim$200 \kms ~larger than that of NGC 3384, $704\pm 2$ \kms.
These results indicate that the three galaxies (M96, M105, and NGC 3384) are close enough to interact with each other. This supports the collisional scenario that the giant gas ring was formed when M96 collided with NGC 3384/M105 \citep{mic10}.

\subsection{The Calibration of the Absolute Magnitudes of SNe Ia and the Hubble Constant}

The distances to M66 and M96 derived in this study can be used to improve the calibration of the absolute magnitudes of SNe Ia. 
Tables 5 and 6 list, respectively, the $V$-band maximum magnitudes of SN 1989B and SN 1998bu derived in this study and previous studies \citep{gib00,san06, tam13}.

Recently \citet{tam13} derived $M_{V,{\rm max}} = -19.45\pm0.15$ for SN 1989B and $M_{V,{\rm max}} = -19.38\pm0.16$ for SN 1998bu from the photometry in the literature \citep{sun99,jha99,her00, wel94}, adopting a mean TRGB distance of the Leo I Group, $(m-M)_0=30.39\pm0.10$.
These values were obtained after correcting for the Galactic extinction, host galaxy extinction, and
decline rates ($\Delta m_{15}$).
These values will become fainter by 0.27 and 0.24 mag if the TRGB distances to M66 and M96 derived in this study are used: $M_{V,{\rm max}} = -19.18\pm0.11$ for SN 1989B and $-19.14\pm0.12$ for SN 1998bu. 
Other previous estimates \citep{gib00,san06} are affected in the similar way, yielding 
$M_{V,{\rm max}} = -19.46\pm0.17$ for SN 1989B and $-19.38\pm0.11$ for SN 1998bu in \citet{gib00}, and
$M_{V,{\rm max}} = -19.17\pm0.06$ for SN 1989B and $-19.11\pm0.06$ for SN 1998bu in \citet{san06}.  


SN 2011fe in M101 is  the nearest recent SN Ia with modern photometry so that it is an excellent object for calibration of SNe Ia.
\citet{lee12} derived maximum magnitudes  of SN 2011fe from the photometry in the literature, adopting a new TRGB distance derived from the weighted mean of nine fields in M101, $M_{V,{\rm max}} = -19.38\pm0.05 ({\rm random})\pm0.12 ({\rm systematic})$.
Thus $V$-band magnitudes of SN 1989B and SN 1998bu are  $\sim0.2$ mag fainter than that of SN 2011fe.
This difference is similar to the dispersion of the absolute magnitudes of SNe Ia, 0.14 \citep{tam13}.
It is noted that the internal extinction for SN 2011fe is known to be negligible, $A_V=0.04$ \citep{pat11}, while those for SN 1989B and 1998bu are not, as listed in Tables 7 and 8, respectively. The values for $A_V$ derived in the previous studies range from $0.82\pm0.08$ to $1.33\pm0.14$ for SN 1989B and from $0.74\pm0.11$ to $1.06\pm0.11$ for SN 1998bu \citep{rei05,wan06b,jha07,tam13}.
Therefore the errors due to internal extinction for SN 1989B and SN 1998bu are expected to be larger than that for SN 2011fe. Further studies to derive better estimates for internal extinction for 
both SNe are needed in the future.

Near-infrared (NIR)  photometry of SN 1998bu in M96 is available in the literature so that SN 1998bu plays as  one of the important calibrators for NIR magnitudes of SNe Ia.   
\citet{tam13} derived $JHK_s$ maximum magnitudes at each band of SN 1998bu from the previous photometry
\citep{jha99, sun99, her00, woo08} : $J=11.55\pm 0.03$, $H=11.59\pm 0.03$, and $K_s=11.42\pm 0.03$.
They adopted a value for internal extinction of $A_V=0.74\pm0.11$. Corresponding extinctions in NIR bands are 
 $A_J = 0.19\pm 0.03$, $A_H = 0.12 \pm 0.02$, and $A_{K_S} = 0.08\pm0.01$.
If we apply internal extinctions presented above 
and adopt our new TRGB distance, we obtain NIR absolute magnitudes of SN 1998bu :  
$M_{J,{\rm max}} = -18.79\pm0.05$, $M_{H,{\rm max}} = -18.68\pm0.05$, and $M_{K_{s},{\rm max}} = -18.81\pm0.04$.
\citet{lee12} derived $JHK_{s}$ magnitudes  of SN 2011fe in M101
from the photometry in \citet{mat12}, adopting a new TRGB distance they derived:
$M_{J,{\rm max}} = -18.79\pm0.04 ({\rm random}) \pm0.12 ({\rm systematic})$, $M_{H,{\rm max}} = -18.55\pm0.04 ({\rm random}) \pm0.12 ({\rm systematic})$, and $M_{K_{s},{\rm max}} = -18.66\pm0.05 ({\rm random}) \pm0.12 ({\rm systematic})$.
Thus absolute $J$ magnitude of SN 1998bu is the same as that of SN 2011fe, while $H, K_s$ magnitudes of SN 1998bu are $\sim0.14$ mag brighter than those of SN 2011fe.
We derive weighted mean values of SN 1989bu and SN 2011fe from these:
$M_{J,{\rm max}} = -18.79\pm0.03$, $M_{H,{\rm max}} = -18.60\pm0.03$, and $M_{K_{s},{\rm max}} = -18.75\pm0.03$. 
It is noted that  these values are $0.2\sim0.4$ mag brighter than recent calibrations of the NIR magnitudes for SNe Ia available in the literature \citep{kri04, fol10, bar12, kat12}.
Recently several calibrations of the NIR absolute magnitudes of SNe Ia were published, but they show a large spread with $\sim 0.2$ mag  differences \citep{woo08,fol10,bur11,man11,kat12,bar12,mat12}. 
Further studies are needed to understand these large differences in the NIR magnitudes of SNe Ia.



The relations between the Hubble constant and the absolute magnitude of SNe Ia are given
by  log $H_0 = 0.2 M_{V,max} + 5 + (0.688 \pm 0.004)$ in \citet{rei05}
or by the equations (2) and (4) in \citet{gib00}. 
Using these relations 
we derive the Hubble constant : 
$H_0 = 69.1 \pm 3.2{\rm (random)}$  \kmsMpc for SN 1989B, 
$H_0 = 71.0 \pm 2.6{\rm (random)}$  \kmsMpc for SN 1998bu, and 
$H_0 = 65.0 \pm 2.1{\rm (random)}$  \kmsMpc for SN 2011fe.
A weighted mean of these three measurement is 
 $H_0 = 67.6 \pm1.5{\rm (random)}  \pm3.7 {\rm (systematic)}$ \kmsMpc. 
Note that this value for the Hubble constant is similar to the recent estimates based on the cosmic microwave background radiation maps in WMAP9 data, $H_0=69.32\pm0.80$ \kmsMpc \citep{ben12} and Planck data $H_0=67.3\pm1.2$ \kmsMpc \citep{ade13}, 
but smaller than other recent determinations based on Cepheid calibration for SNe Ia luminosity, 
$H_0 = 74 \pm3$ km s$^{-1}$ Mpc$^{-1}$  \citep{rie11, fre12} .

\begin{table*}[htp]
\centering
\scriptsize 
\caption{A Summary of Optical Luminosity Calibrations for SN 1989B in M66}
\begin{tabular}{c l c c c c}

\hline \hline
ID 	& References 	& $(m-M)_0$ & $V_{corr}$$^a$& $M_{V}$$^a$ & $H_{0} [km/s/Mpc]$ \\
\hline
(1) & \citet{tam13}  & $30.39\pm0.10$ & $10.94\pm0.11$ & $-19.45\pm0.15$ & $62.8\pm4.3$ \\
(2) & \citet{gib00}	 & $30.06\pm0.17$ & $10.66\pm0.16^b$ & $-19.40\pm0.24^b$  & $68.9\pm7.5$\\
(3) & \citet{san06}  & $30.50\pm0.10$ & $10.95\pm0.05$ & $-19.55\pm0.11$ & $60.0\pm3.1$ \\
\hline
(4) & This study	& $30.12\pm0.03$ &\multicolumn{1}{l}{\citet{tam13}}& $-19.18\pm0.11$ & $71.1\pm3.8$ \\
(5) & This study	& $30.12\pm0.03$ &\multicolumn{1}{l}{\citet{gib00}}& $-19.46\pm0.17^b$ & $67.1\pm5.2$ \\
(6) & This study	& $30.12\pm0.03$ &\multicolumn{1}{l}{\citet{san06}}& $-19.17\pm0.06$ & $71.4\pm2.0$ \\
 \hline
(7) & This study	& $30.12\pm0.03$ &\multicolumn{2}{l}{Straight mean of (4) and (5) }	& $69.1\pm3.2$\\
(8) & This study	& $30.12\pm0.03$ &\multicolumn{2}{l}{Weighted mean of (4) and (5) }	& $69.7\pm3.1$\\
\hline
\multicolumn{6}{l}{%
  \begin{minipage}{14cm}%
    $^a$ Corrected for the Galactic extinction, host galaxy extinction and decline rate ($\Delta m_{15}$).\\
    $^b$ We applied decline rate ($\Delta m_{15}$) correction using equation (21) of \citet{phi99}.\\
  \end{minipage}%
}\\
\end{tabular}
\end{table*}


\begin{table*}[htp]
\centering
\scriptsize
\caption{A Summary of Optical Luminosity Calibrations for SN 1998bu in M96}
\begin{tabular}{c l c c c c}
\hline \hline
ID & References & $(m-M)_0$ & $V_{corr}$$^a$ & $M_{V}$$^a$  & $H_{0} [km/s/Mpc]$ \\
\hline
(1) & \citet{tam13} & $30.39\pm0.10$ & $11.01\pm0.12$ & $-19.38\pm0.16$ & $64.9\pm4.7$\\
(2) & \citet{gib00} & $30.20\pm0.10$ & $10.77\pm0.11^b$ & $-19.43\pm0.15^b$ & $68.0\pm4.7$\\
(3) & \citet{san06} & $30.34\pm0.11$ & $11.04\pm0.05$ & $-19.30\pm0.12$ & $67.3\pm3.8$\\ 
\hline
(4) & This study	& $30.15\pm0.03$ &\multicolumn{1}{l}{ \citet{tam13}} & $-19.14\pm0.12$ & $72.4\pm3.9$\\
(5) & This study	& $30.15\pm0.03$ &\multicolumn{1}{l}{ \citet{gib00}} & $-19.38\pm0.11^b$ & $69.6\pm3.6$\\
(6) & This study	& $30.15\pm0.03$ &\multicolumn{1}{l}{\citet{san06}}  & $-19.11\pm0.06$ & $73.5\pm2.1$\\
 \hline
(7) & This study	& $30.15\pm0.03$ &\multicolumn{2}{l}{Straight mean of (4) and (5)}	& $71.0\pm2.6$\\
(8) & This study	& $30.15\pm0.03$ &\multicolumn{2}{l}{Weighted mean of (4) and (5)}	& $70.9\pm2.6$\\
\hline
\multicolumn{6}{l}{%
  \begin{minipage}{14cm}%
    $^a$ Same as Table 5.\\
    $^b$ Same as Table 5.
  \end{minipage}%
}\\
\end{tabular}
\end{table*}

\begin{table*}[htp]
\centering
\small
\caption{A Summary of Internal Extinction Values for SN 1989B in M66}
\begin{tabular}{l c c c  c}
\hline \hline
References & $E(B-V)$ & $A_V$ & $R_V$  & Remarks \\
\hline
\citet{sun99}   & $0.37\pm0.03$ & $1.15\pm0.09$ 	& $3.1$  & $BVI$\\
\citet{jha07}   & $0.47\pm0.07$ & $1.33\pm0.14$ 	& $2.86\pm0.29$  & $UBVRI$\\
\citet{wan06b}  & $0.42\pm0.06$ & $0.97\pm0.14$		& $2.30\pm0.11$  & $UBVI$\\
\citet{rei05} $\&$& $0.31\pm0.03$ & $0.82\pm0.08$ 	& $2.65\pm0.15$  & $BVI$ \\
\citet{tam13}	& 				&					&				 & \\
\hline
\end{tabular}
\end{table*}

%
\begin{table*}[htp]
\centering
\small
\caption{A Summary of Internal Extinction Values for SN 1998bu in M96}
\begin{tabular}{l c c c  c }
\hline \hline
References & $E(B-V)$ & $A_V$ & $R_V$ & Remarks \\
\hline
\citet{sun99}   & $0.34\pm0.03$ & $1.05\pm0.09$ 	& $3.1$  & $BVI$\\
\citet{jha07}   & $0.34\pm0.05$ & $1.06\pm0.11$ 	& $3.13\pm0.36$  & $UBVRI$\\
\citet{wan06b}  & $0.36\pm0.04$ & $0.83\pm0.10$		& $2.30\pm0.11$  & $UBVI$\\
\citet{rei05} $\&$& $0.28\pm0.04$ & $0.74\pm0.11$ 	& $2.65\pm0.15$  & $BVI$ \\
\citet{tam13}	& 				&					&				 & \\
\hline
\end{tabular}
\end{table*}

\section{Summary}

We present $VI$ photometry of the resolved stars in two spiral galaxies M66 and M96 that host  SNe Ia in the Leo I Group, derived from  $HST$/ACS $F555W$ and $F814W$ images. 
Then we estimate the distances to these two galaxies applying the TRGB method to this photometry.  
We summarize main results in the following.

\begin{itemize}
\item Most of the resolved stars in the selected regions of M66 and M96 are red giants,  allowing us to determine the distances to these galaxies. 

\item 
The $I$-band magnitudes of the TRGB are found to be 
$I_{\rm TRGB}=26.20\pm0.03$ for M66 and $26.21\pm0.03$ for M96. 
These TRGB magnitudes yield distance modulus 
$(m-M)_0=30.12\pm0.03({\rm random}) \pm 0.12 ({\rm systematic})$ for M66 and
$(m-M)_0=30.15\pm0.03({\rm random}) \pm 0.12 ({\rm systematic})$ for M96. 
 This result shows that M66 and M96 are the members
 of the same group.

\item The absolute maximum magnitudes of the  SNe Ia 
are derived from the previous photometry and the distance measurement in this study, as listed in Tables 5 and 6.
Similarly we derive NIR magnitudes for SN 1998bu: $M_{J, {\rm max}} = -18.79\pm0.05$, $M_{H,{\rm max}} = -18.68\pm0.05$, and $M_{K_{s}, {\rm max}} = -18.81\pm0.04$.

\item 
Combining the results for SN 1989B and SN 1998bu with those for SN 2011fe in M101 based on the same method given in \citet{lee12}, 
we obtain an estimate of the Hubble constant, $H_0=67.6 \pm1.5 \pm3.7 $ \kmsMpc.

\end{itemize}

\bigskip

This paper is based on image data obtained from the Multimission Archive at the Space Telescope Science Institute (MAST).
The authors would like to thank Won-Kee Park for technical support in image processing. 
This work was supported by the National Research Foundation of Korea (NRF) grant
funded by the Korea Government (MEST) (No. 2012R1A4A1028713).



\clearpage


\begin{thebibliography}{}



\bibitem[Ajhar et al.(2001)]{ajh01} Ajhar, E.~A., Tonry, 
J.~L., Blakeslee, J.~P., Riess, A.~G., 
\& Schmidt, B.~P.\ 2001, \apj, 559, 584 

\bibitem[Barone-Nugent et al.(2012)]{bar12} Barone-Nugent, R.~L., Lidman, C., Wyithe, J.~S.~B., et al.\ 2012, \mnras, 425, 1007 
9
\bibitem[Bellazzini et al.(2004)]{bel04} Bellazzini, M., Ferraro, F.~R., Sollima, A., Pancino, E., \& Origlia, L.\ 2004, \aap, 424, 199 

\bibitem[Bennett et al.(2012)]{ben12} Bennett, C.~L., Larson, D., Weiland, J.~L., et al.\ 2012, arXiv:1212.5225 


\bibitem[Burns et al.(2011)]{bur11} Burns, C.~R., Stritzinger, M., Phillips, M.~M., et al.\ 2011, \aj, 141, 19 



\bibitem[Ciardullo et al.(2002)]{cia02} Ciardullo, R., 
Feldmeier, J.~J., Jacoby, G.~H., et al.\ 2002, \apj, 577, 31

\bibitem[Ciatti \& Rosino(1977)]{cia77} Ciatti, F., \& Rosino, L.\ 1977, \aap, 56, 59 


\bibitem[de Vaucouleurs(1975)]{dev75}  de Vaucouleurs, G. 1975, in Stars and Stellar Systems 9 : Galaxies and the Universe, ed. A. Sandage, M. Sandage, \& J. Kristian (Chicago : Univ.
Chicago Press), 557



\bibitem[Dolphin 
\& Kennicutt(2002)]{dol02} Dolphin, A.~E., \& Kennicutt, R.~C., Jr.\ 2002, \aj, 123, 207

\bibitem[Elias-Rosa et al.(2011)]{eli11} Elias-Rosa, N., Van Dyk, S.~D., Li, W., et al.\ 2011, \apj, 742, 

\bibitem[Evans \& McNaught(1989)]{eva89} Evans, R.~O., \& McNaught, R.~H.\ 1989, \iaucirc, 4726, 1 

\bibitem[Feldmeier et al.(1997)]{fel97} Feldmeier, J.~J., 
Ciardullo, R., \& Jacoby, G.~H.\ 1997, \apj, 479, 231

\bibitem[Folatelli et al.(2010)]{fol10} Folatelli, G., Phillips, M.~M., Burns, C.~R., et al.\ 2010, \aj, 139, 120 

\bibitem[Freedman et al.(2001)]{fre01} Freedman, W.~L., 
Madore, B.~F., Gibson, B.~K., et al.\ 2001, \apj, 553, 47

\bibitem[Freedman \& Madore(2010)]{fre10} Freedman, W.~L., \& Madore, B.~F.\ 2010, \araa, 48, 673 

\bibitem[Freedman et al.(2012)]{fre12} Freedman, W.~L., Madore, B.~F., Scowcroft, V., et al.\ 2012, \apj, 758, 24 

\bibitem[Gibson et al.(2000)]{gib00} Gibson, B.~K., Stetson, 
P.~B., Freedman, W.~L., et al.\ 2000, \apj, 529, 723  

\bibitem[Gibson 
\& Stetson(2001)]{gib01} Gibson, B.~K., \& Stetson, P.~B.\ 2001, \apjl, 547, L103 

R.~L., Freedman, W.~L., et al.\ 1997, \apj, 477, 535  


\bibitem[Harris et al.(2007a)]{har07a} Harris, W.~E., Harris, 
G.~L.~H., Layden, A.~C., \& Stetson, P.~B.\ 2007, \aj, 134, 43  


\bibitem[Harris et al.(2007b)]{har07b} Harris, W.~E., Harris, 
G.~L.~H., Layden, A.~C., \& Wehner, E.~M.~H.\ 2007, \apj, 666, 903  


\bibitem[Hernandez et al.(2000)]{her00} Hernandez, M., 
Meikle, W.~P.~S., Aparicio, A., et al.\ 2000, \mnras, 319, 223  


\bibitem[Hislop et al.(2011)]{his11} Hislop, L., Mould, J., 
Schmidt, B., et al.\ 2011, \apj, 733, 75  


\bibitem[Jang et al.(2012)]{jan12} Jang, I.~S., Lim, S., 
Park, H.~S., \& Lee, M.~G.\ 2012, \apjl, 751, L19

\bibitem[Jensen et al.(2003)]{jen03} Jensen, J.~B., Tonry, 
J.~L., Barris, B.~J., et al.\ 2003, \apj, 583, 712 


\bibitem[Jha et al.(1999)]{jha99} Jha, S., Garnavich, P.~M., 
Kirshner, R.~P., et al.\ 1999, \apjs, 125, 73  


\bibitem[Jha et al.(2007)]{jha07} Jha, S., Riess, A.~G., 
\& Kirshner, R.~P.\ 2007, \apj, 659, 122  


\bibitem[Kanbur et 
al.(2003)]{kan03} Kanbur, S.~M., Ngeow, C., Nikolaev, S., Tanvir, N.~R., \& Hendry, M.~A.\ 2003, \aap, 411, 361



\bibitem[Kattner et al.(2012)]{kat12} Kattner, S., Leonard, 
D.~C., Burns, C.~R., et al.\ 2012, \pasp, 124, 114  


\bibitem[Kelson et al.(2000)]{kel00} Kelson, D.~D., 
Illingworth, G.~D., Tonry, J.~L., et al.\ 2000, \apj, 529, 768  

\bibitem[Kochanek(1997)]{koc97} Kochanek, C.~S.\ 1997, \apj, 
491, 13

\bibitem[Krisciunas et al.(2004)]{kri04} Krisciunas, K., 
Phillips, M.~M., \& Suntzeff, N.~B.\ 2004, \apjl, 602, L81  


\bibitem[Lee et al.(1993)]{lee93} Lee, M.~G., Freedman, 
W.~L., \& Madore, B.~F.\ 1993, \apj, 417, 553  


\bibitem[Lee 
\& Jang(2012)]{lee12} Lee, M.~G., \& Jang, I.~S.\ 2012, \apjl, 760, L14 (Paper I) 









\bibitem[Mandel et al.(2009)]{man09} Mandel, K.~S., 
Wood-Vasey, W.~M., Friedman, A.~S., 
\& Kirshner, R.~P.\ 2009, \apj, 704, 629

\bibitem[Mandel et al.(2011)]{man11} Mandel, K.~S., Narayan, 
G., \& Kirshner, R.~P.\ 2011, \apj, 731, 120  


\bibitem[Matheson et al.(2012)]{mat12} Matheson, T., Joyce, 
R.~R., Allen, L.~E., et al.\ 2012, \apj, 754, 19  




\bibitem[M{\'e}ndez et al.(2002)]{men02} M{\'e}ndez, B., 
Davis, M., Moustakas, J., et al.\ 2002, \aj, 124, 213  


\bibitem[Michel-Dansac et al.(2010)]{mic10} Michel-Dansac, 
L., Duc, P.-A., Bournaud, F., et al.\ 2010, \apjl, 717, L143  


\bibitem[Mouhcine et al.(2010)]{mou10} Mouhcine, M., Harris, 
W.~E., Ibata, R., \& Rejkuba, M.\ 2010, \mnras, 404, 1157  




\bibitem[Mould 
\& Sakai(2009a)]{mou09a} Mould, J., \& Sakai, S.\ 2009a, \apj, 694, 1331  



\bibitem[Mould 
\& Sakai(2009b)]{mou09b} Mould, J., \& Sakai, S.\ 2009b, \apj, 697, 996  



\bibitem[Mueller 
\& Hoeflich(1994)]{mue94} Mueller, E., \& Hoeflich, P.\ 1994, \aap, 281, 51




\bibitem[Nowak et al.(2010)]{now10} Nowak, N., Thomas, J., 
Erwin, P., et al.\ 2010, \mnras, 403, 646  

\bibitem[Patat et al.(2011)]{pat11} Patat, F., Cordiner, 
M.~A., Cox, N.~L.~J., et al.\ 2011, arXiv:1112.0247 

\bibitem[Paturel et 
al.(2002)]{pat02} Paturel, G., Teerikorpi, P., Theureau, G., et al.\ 2002, \aap, 389, 19 

\bibitem[Phillips et al.(1999)]{phi99} Phillips, M.~M., Lira, 
P., Suntzeff, N.~B., et al.\ 1999, \aj, 118, 1766 


\bibitem[Pierce(1994)]{pie94} Pierce, M.~J.\ 1994, \apj, 430, 
53 


\bibitem[Planck Collaboration et al.(2013)]{ade13} Planck 
Collaboration, Ade, P.~A.~R., Aghanim, N., et al.\ 2013, arXiv:1303.5076


\bibitem[Reindl et al.(2005)]{rei05} Reindl, B., Tammann, 
G.~A., Sandage, A., \& Saha, A.\ 2005, \apj, 624, 532  




\bibitem[Riess et al.(2011)]{rie11} Riess, A.~G., Macri, L., 
Casertano, S., et al.\ 2011, \apj, 730, 119  


\bibitem[Rizzi et al.(2007)]{riz07} Rizzi, L., Tully, R.~B., 
Makarov, D., et al.\ 2007, \apj, 661, 815  

\bibitem[Russell(2002)]{rus02} Russell, D.~G.\ 2002, \apj, 
565, 681 

\bibitem[Saha et al.(1999)]{sah99} Saha, A., Sandage, A., 
Tammann, G.~A., et al.\ 1999, \apj, 522, 802  


\bibitem[Saha et al.(2006)]{sah06} Saha, A., Thim, F., 
Tammann, G.~A., Reindl, B., \& Sandage, A.\ 2006, \apjs, 165, 108  


\bibitem[Sakai et al.(1996)]{sak96} Sakai, S., Madore, B.~F., 
\& Freedman, W.~L.\ 1996, \apj, 461, 713  






\bibitem[Sandage et al.(2006)]{san06} Sandage, A., Tammann, 
G.~A., Saha, A., et al.\ 2006, \apj, 653, 843  


\bibitem[Schlafly 
\& Finkbeiner(2011)]{sch11} Schlafly, E.~F., \& Finkbeiner, D.~P.\ 2011, \apj, 737, 103  


\bibitem[Schlegel et al.(1998)]{sch98} Schlegel, D.~J., 
Finkbeiner, D.~P., \& Davis, M.\ 1998, \apj, 500, 525  


\bibitem[Schneider et al.(1983)]{sch83} Schneider, S.~E., 
Helou, G., Salpeter, E.~E., \& Terzian, Y.\ 1983, \apjl, 273, L1  


\bibitem[Schneider(1989)]{sch89} Schneider, S.~E.\ 1989, 
\apj, 343, 94  






\bibitem[Sirianni et al.(2005)]{sir05} Sirianni, M., Jee, 
M.~J., Ben{\'{\i}}tez, N., et al.\ 2005, \pasp, 117, 1049  



\bibitem[Springob et al.(2009)]{spr09} Springob, C.~M., 
Masters, K.~L., Haynes, M.~P., Giovanelli, R., 
\& Marinoni, C.\ 2009, \apjs, 182, 474


\bibitem[Spyromilio et 
al.(2004)]{spy04} Spyromilio, J., Gilmozzi, R., Sollerman, J., et al.\ 2004, \aap, 426, 547  




\bibitem[Stetson(1994)]{ste94} Stetson, P.~B.\ 1994, \pasp, 
106, 250  




\bibitem[Suntzeff et al.(1999)]{sun99} Suntzeff, N.~B., 
Phillips, M.~M., Covarrubias, R., et al.\ 1999, \aj, 117, 1175  



\bibitem[Tammann 
\& Reindl(2013)]{tam13} Tammann, G.~A., \& Reindl, B.\ 
2013, \aap, 549, 136

\bibitem[Takanashi et al.(2008)]{tak08} Takanashi, N., Doi, 
M., \& Yasuda, N.\ 2008, \mnras, 389, 1577

\bibitem[Tanvir et al.(1995)]{tan95} Tanvir, N.~R., Shanks, 
T., Ferguson, H.~C., \& Robinson, D.~R.~T.\ 1995, \nat, 377, 27  


\bibitem[Tanvir et al.(1999)]{tan99} Tanvir, N.~R., Ferguson, 
H.~C., \& Shanks, T.\ 1999, \mnras, 310, 175  


\bibitem[Thilker et al.(2009)]{thi09} Thilker, D.~A., 
Donovan, J., Schiminovich, D., et al.\ 2009, \nat, 457, 990  

\bibitem[Tonry et al.(2001)]{ton01} Tonry, J.~L., Dressler, 
A., Blakeslee, J.~P., et al.\ 2001, \apj, 546, 681

\bibitem[Tully et al.(2009)]{tul09} Tully, R.~B., Rizzi, L., 
Shaya, E.~J., et al.\ 2009, \aj, 138, 323  


\bibitem[Van Dyk et al.(2000)]{van00} Van Dyk, S.~D., Peng, 
C.~Y., King, J.~Y., et al.\ 2000, \pasp, 112, 1532  


\bibitem[Villi et al.(1998)]{vil98} Villi, M., Nakano, S., 
Aoki, M., Skiff, B.~A., \& Hanzl, D.\ 1998, \iaucirc, 6899, 1  






\bibitem[Wang et al.(2006)]{wan06b} Wang, X., Wang, L., Pain, 
R., Zhou, X., \& Li, Z.\ 2006, \apj, 645, 488   


\bibitem[Wells et al.(1994)]{wel94} Wells, L.~A., Phillips, 
M.~M., Suntzeff, B., et al.\ 1994, \aj, 108, 2233 


\bibitem[Willick 
\& Batra(2001)]{wil01} Willick, J.~A., \& Batra, P.\ 2001, \apj, 548, 564  


\bibitem[Wood-Vasey et al.(2008)]{woo08} Wood-Vasey, W.~M., 
Friedman, A.~S., Bloom, J.~S., et al.\ 2008, \apj, 689, 377  
\end{thebibliography}
\end{document}